\documentclass[sigconf]{acmart}
\usepackage{multirow}
\usepackage{balance} 
\usepackage{colortbl}
\usepackage{amsmath}
\usepackage{enumitem}
\usepackage{algorithm}
\usepackage{algpseudocode}
\usepackage{float}
\usepackage{graphicx}
\usepackage{array}
\usepackage{subcaption}
\usepackage[normalem]{ulem}
\useunder{\uline}{\ul}{}
\algnewcommand\algorithmicforeach{\textbf{for each}}
\algdef{S}[FOR]{ForEach}[1]{\algorithmicforeach\ #1\ \algorithmicdo}
\newcolumntype{V}[1]{>{\centering\arraybackslash\rotatebox{90}}m{#1}}
\AtBeginDocument{%
  }

\copyrightyear{2026}
\acmYear{2026}
\setcopyright{cc}
\setcctype{by}
\acmConference[SIGIR '26]{Proceedings of the 49th International ACM SIGIR Conference on Research and Development in Information Retrieval}{July 20--24, 2026}{Melbourne, VIC, Australia}
\acmBooktitle{Proceedings of the 49th International ACM SIGIR Conference on Research and Development in Information Retrieval (SIGIR '26), July 20--24, 2026, Melbourne, VIC, Australia}
\acmDOI{10.1145/3805712.3809691}
\acmISBN{979-8-4007-2599-9/2026/07}

\copyrightyear{2026}
\acmYear{2026}
\setcopyright{cc}
\setcctype{by}
\acmConference[SIGIR '26]{Proceedings of the 49th International ACM SIGIR Conference on Research and Development in Information Retrieval}{July 20--24, 2026}{Melbourne, VIC, Australia}
\acmBooktitle{Proceedings of the 49th International ACM SIGIR Conference on Research and Development in Information Retrieval (SIGIR '26), July 20--24, 2026, Melbourne, VIC, Australia}
\acmDOI{10.1145/3805712.3809691}
\acmISBN{979-8-4007-2599-9/2026/07}

\begin{document}

\title{Prompt-Unknown Promotion Attacks against LLM-based Sequential Recommender Systems}

\author{Yuchuan Zhao}
\affiliation{%
  \institution{The University of Queensland}
  \city{Brisbane}
  \country{Australia}
}
\email{y.c.zhao@uq.edu.au}

\author{Tong Chen}
\affiliation{%
  \institution{The University of Queensland}
  \city{Brisbane}
  \country{Australia}
}
\email{tong.chen@uq.edu.au}

\author{Junliang Yu}
\affiliation{%
  \institution{Griffith University}
  \city{Brisbane}
  \country{Australia}
}
\email{junliang.yu@griffith.edu.au}

\author{Zongwei Wang}
\affiliation{%
  \institution{Chongqing University}
  \city{Chongqing}
  \country{China}
}
\email{zongwei@cqu.edu.cn}

\author{Lizhen Cui}
\affiliation{%
  \institution{Shandong University}
  \city{Jinan}
  \country{China}}
\email{clz@sdu.edu.cn}

\author{Hongzhi Yin}
\authornote{Corresponding author}
\affiliation{%
  \institution{The University of Queensland}
  \city{Brisbane}
  \country{Australia}
}
\email{h.yin1@uq.edu.au}

\renewcommand{\shortauthors}{Yuchuan Zhao et al.}

\begin{abstract}
Large language model-powered sequential recommender systems (LLM-SRSs) have recently demonstrated remarkable performance, enabling recommendations through prompt-driven inference over user interaction sequences. However, this paradigm also introduces new security vulnerabilities, particularly text-level manipulations, rendering them appealing targets for promotion attacks that purposely boost the ranking of specific target items. Although such security risks have been receiving increasing attention, existing studies typically rely on an unrealistic assumption of access to either the victim model or prompt to unveil attack mechanisms. In this work, we investigate the item promotion attack in LLM-SRSs under a more realistic setting where both the system prompt and victim model are unknown to the attacker, and propose a \textbf{P}rompt-\textbf{U}nknown \textbf{D}ual-poisoning \textbf{A}ttack (PUDA) framework. To simulate attacks under this full black-box setting, we introduce an LLM-based evolutionary refinement strategy that infers discrete system prompts, enabling the training of an effective surrogate model that mimics the behaviors of the victim model. Leveraging the distilled prompt and surrogate model, we devise a promotion attack that adversarially revises target item texts under semantic constraints, which is further complemented by the highly plausible, surrogate-generated poisoning sequences to enable cost-effective target item promotion. Extensive experiments on real-world datasets demonstrate that PUDA consistently outperforms state-of-the-art competitors in boosting the exposure of unpopular target items. Our findings reveal critical security risks in modern LLM-SRSs even when both prompts and models are protected, and highlight the need for more robust defensive means.
\end{abstract}



\maketitle

\section{Introduction}
Sequential recommender systems (SRSs) underpin modern digital platforms by modeling users’ evolving interests and item transition dynamics, enabling accurate next-item prediction and personalized recommendation at scale \cite{kang2018self, xie2022contrastive, yin2025device}. Recently, the emergence of large language models (LLMs) has opened new opportunities for advancing SRSs by enabling richer semantic modeling and more expressive representations of user behavior and item content \cite{harte2023leveraging,zheng2024harnessing}. Leveraging their strong language understanding and in-context learning capabilities \cite{wu2024survey,cao2024aligning}, LLMs give rise to a novel recommendation paradigm that verbalizes user–item interactions as natural language prompts. Representative approaches, such as P5 \cite{geng2022recommendation}, TALLRec \cite{bao2023tallrec}, and SPRec \cite{gao2025sprec}, demonstrate that LLMs can effectively interpret these prompts to capture complex interaction patterns, leading to improved recommendation performance.

Due to their central role in industrial applications, SRSs have increasingly become prominent targets for malicious attacks \cite{nguyen2024manipulating}. To unveil the security loopholes of SRSs, prior studies have explored a wide range of adversarial strategies against traditional SRSs, including poisoning attacks \cite{wang2023poisoning,zhao2025diversity}, model extraction attacks \cite{wang2025sim4rec,yue2021black}, and membership inference attacks \cite{zhu2023membership,he2025membership,zhang2021membership}. Among these threats, promotion attacks \cite{zhang2022pipattack,chen2024adversarial,yuan2023manipulating} which aim at artificially increasing the visibility of specific target items by injecting manipulated interactions or crafting adversarial content, have emerged as the most prominent and widely studied attack type in SRSs.

With the recent adoption of LLMs, a natural question arises as to whether LLM-powered SRSs (LLM-SRSs) inherit similar vulnerabilities to promotion attacks \cite{zhang2024lorec}. Intuitively, given LLM-SRSs' strong reliance on their textual inputs, i.e., prompt instructions and item contexts, adversaries can attempt to manipulate recommendation results via semantic perturbations. Several recent studies support this intuition: CheatAgent \cite{ning2024cheatagent} adversarially modifies the original prompt to degrade system performance. Meanwhile, TextSimu \cite{wang2025id} and Zhang et al. \cite{zhang2024stealthy} focus on item-side textual manipulations, rewriting target item descriptions to mimic popular items and thereby increase their visibility. 

However, these methods are built upon unrealistic assumptions that are unlikely to hold in real-world deployments. In practice, both the LLM-SRS backbone and the system prompt are strictly proprietary and inaccessible to attackers. As a result, prior studies \cite{ning2024cheatagent,he2025membership} that rely on direct access to the system prompt are fundamentally ill-posed. Under prompt-unknown constraint, a natural strategy is to distill the discrete system prompt to approximate the original one. Nevertheless, this direction remains largely underexplored in LLM-SRSs, leaving a critical gap in understanding how attacks can be conducted under practical black-box constraints. 
Moreover, existing methods \cite{wang2025id,zhang2024stealthy} predominantly rely on modifying item-side textual attributes (e.g., titles or descriptions) to promote target items, typically by imitating the textual characteristics of high-popularity items. Such text-centric heuristics often distort original item semantics or even introduce misleading descriptions, making the manipulated items easily detectable by users or platform moderation mechanisms in real-world deployments. More importantly, such attacks mainly focus on text-level manipulation and overlook the sequential dependencies and item co-occurrence patterns inherent in user interaction sequences, thus failing to generate realistic and effective poisoning signals at the interaction level.

To investigate the security risks of LLM-SRSs under a more realistic setting, we propose a \textbf{P}rompt-\textbf{U}nknown \textbf{D}ual-poisoning \textbf{A}ttack framework (\textbf{PUDA}) against LLM-SRSs. PUDA framework operates under black-box assumptions and integrates two tightly coupled components. Firstly, we introduce an LLM-driven evolutionary prompt distillation strategy that employs crossover and mutation \cite{fernando2023promptbreeder,cui2025see} to infer high-fidelity discrete prompts without accessing proprietary instructions, while simultaneously generating plausible interaction histories to train a surrogate recommender that mimics the victim black-box model. Secondly, building upon the distilled prompt and the surrogate model, we design a dual-poisoning item promotion attack strategy that jointly refines target item's textual descriptions and injects behavior-preserving malicious user sequences, thereby amplifying item exposure while maintaining consistency with real user behavior. The design of PUDA indicates that LLM-SRSs, despite operating in inaccessible black-box settings, remain amenable to prompt distillation and thus constitute vulnerable targets for stealthy item-promotion attacks.

The main contributions of this paper are summarized as follows:

\begin{itemize}[leftmargin=*]
    \item We propose a Prompt-Unknown Dual-Poisoning Attack  framework for LLM-SRSs under a realistic setting where model internals and prompts are entirely inaccessible. 
    \item We design an LLM-driven evolutionary prompt distillation strategy that efficiently infers high-fidelity discrete prompts for SRS tasks, enabling accurate surrogate model construction without access to proprietary system instructions.
    \item We develop a dual-poisoning attack strategy that jointly performs controlled semantic refinement of target item texts and behavior-preserving poisoning sequence injection. This coordinated design enables effective long-tail item promotion in black-box LLM-SRSs under limited attacker budgets.
    \item Extensive experiments demonstrate that our method substantially improves target item exposure in black-box settings, revealing significant security vulnerabilities in modern LLM-SRSs.
\end{itemize}

\section{Related Work}
\subsection{Promotion Attacks against SRSs}

Most existing promotion attacks against traditional SRSs aim to manipulate recommendation outcomes by injecting fake user interactions or fabricated user profiles. LOKI \cite{zhang2020practical} constructs a local simulator to imitate the target model and employs reinforcement learning (RL) to train the attack agent that generates adversarial sequences. Wang et al. \cite{wang2023poisoning} leverage generative adversarial networks (GANs) to synthesize fake sequences without requiring API access or prior knowledge of the target model. DDSP \cite{zhao2025diversity} develops a diversity-aware poisoning sequence generator with a revamped attack objective to promote target items more effectively. 

With the emergence of LLMs, new perspectives have been introduced for designing promotion attack strategies. TextSimu \cite{wang2025id} employs LLMs to rewrite target items' descriptions to mimic popular items, while Agent4SR \cite{gu2025llm} injects fake user profiles generated by LLM-based agents to manipulate recommendation behaviors. More recently, the security of LLM-SRSs under adversarial settings has begun to attract increasing attention. Zhang et al. \cite{zhang2024stealthy} propose a testing-phase promotion attack that leverages LLMs to rewrite the textual description of the target item, revealing the vulnerability of LLM-SRSs to semantic manipulation. CheatAgent \cite{ning2024cheatagent} introduces LLM agents to generate adversarial perturbations injected into the system prompt, aiming to disrupt the decision-making process of SRSs. To the best of our knowledge, no prior work has investigated promotion attacks under a black-box setting where both the LLM-SRS and its underlying prompt are unknown.

\subsection{Prompt Distillation on LLM-SRSs}
In LLM-based recommendation, prompts critically influence recommendation quality but are typically private and inaccessible. Prompt distillation is therefore essential for approximating black-box recommenders. POD \cite{li2023prompt} distills discrete prompts into continuous soft prompt vectors to bridge item IDs and textual semantics, thus reducing inference cost. RLPrompt \cite{deng2022rlprompt} formulates discrete prompt search as a reinforcement learning problem by maximizing downstream task performance. CoPrompt \cite{cho2023discrete} further proposes a left-to-right discrete prompt optimization at the cost of high computational overhead due to the large search space.



Given the strong reasoning ability of LLMs, several studies have explored using LLMs to distill and optimize prompts. APE \cite{zhou2022large} leverages LLMs to automatically generate and select instruction candidates based on input-output demonstrations. PromptBreeder \cite{fernando2023promptbreeder} introduces a self-referential evolutionary framework that iteratively adapts prompts for specific domains, while SEE \cite{cui2024see} proposes a cohesive in-context prompt optimization method that jointly refines instructions and examples via metaheuristic search. Despite these advances, discrete prompt distillation for recommendation tasks remains under-explored. 


\section{Preliminaries}
\subsection{LLM-based Sequential Recommendation}\label{sec:SRSs}
A sequential recommender system (SRS) contains a user set \(\mathcal{U}\) and an item set \(\mathcal{I}\). Each user \(u \in \mathcal{U}\) is associated with a chronological interaction sequence \(s_u = \{i^{(u)}_1,i^{(u)}_2,...,i^{(u)}_M\}\), where $M$ denotes the maximum sequence length and $|\mathcal{U}| = N$ is the number of users. The goal of sequential recommendation \cite{kang2018self} is to predict the next item that a user $u$ is most likely to interact with: $i_u^* = \arg \max _{i \in \mathcal{I}} p\left(i_{M+1}^{(u)}=i \mid s_u\right).$ In LLM-SRSs, this prediction task is reformulated as a language modeling problem \cite{bao2023large} by converting user interaction histories into natural language prompts. Without loss of generality, we define a sequential recommendation prompt $\textit{P}$ as the complete textual input to the LLM, typically consisting of:
\begin{itemize}[leftmargin=*]
    \item \textbf{Instruction $I$}: A natural-language description specifying the next-item prediction task based on a given input sequence.
    \item \textbf{Demonstrations (Demos) $\mathcal{X}$}: Example input-output pairs illustrating how the model should map an interaction sequence to its next-item label. Each demo includes (i) a historical sequence $s_u$, and (ii) its ground-truth next item $y_u$.
\end{itemize}
Formally, a prompt is defined as: $P=\left[I, \mathcal{X}_u\right]$, where $I$ denotes the natural language instruction and $\mathcal{X}_u=(s_u, y_u)$ is a demonstration pair consisting of a user's historical interaction sequence $s_u$ and the corresponding next-item label $y_u$. In the following, we use $P$ to denote an individual prompt and $\mathcal{P}$ to represent a prompt set. LLM-SRSs infer user preferences by jointly interpreting the semantics of the instruction and the interaction context. Unless otherwise specified, we use the term \emph{prompt} $P$ interchangeably with \emph{instruction} $I$,
and retain the term \emph{prompt distillation} to remain consistent with prior work \cite{deng2022rlprompt}.

\subsection{Attacker Brief}
\label{Attacker Brief}
Building on the LLM-SRSs framework described above, we characterize the adversary from three perspectives.

\noindent\textbf{Attacker's Goal.}
Poisoning attacks in recommender systems can be broadly categorized into two types: (i) non-targeted attacks, which aim to degrade overall performance \cite{ning2024cheatagent,yi2023ua}, and (ii) targeted attacks, which seek to promote or demote specific items \cite{nguyen2024manipulating,yuan2025robust}. In this work, we focus on targeted promotion attacks \cite{wang2025id, wang2024unveiling}, where the adversary attempts to increase the exposure of target items across users' recommendation list in an LLM-SRS.

\noindent\textbf{Attacker's Knowledge.}
We consider a realistic black-box setting \cite{song2020poisonrec,chen2022knowledge} for attacking LLM-SRSs. The attacker has no access to the victim model's architecture or parameters. Furthermore, the system prompt which is proprietary to the platform, cannot be accessed or edited. While the victim model is trained on all interactions $\mathcal{D}$, the attacker can only observe partial interaction data denoted by $\mathcal{D}'\subseteq \mathcal{D}$. This setting replicates the practical e-commerce environment where only a fraction of the interactions are made public, e.g., not all Amazon users publish a product review after the purchase.

\noindent\textbf{Attacker's Capability.}
The attacker is able to create a small number of fake accounts $\mathcal{U}_F$, and inject forged interactions into the platform. In addition, attackers are allowed to issue a limited number of queries to the black-box victim model to obtain corresponding recommendation results, but cannot access internal information such as token probabilities, model parameters, or gradients.

\begin{figure*}
    \setlength{\abovecaptionskip}{0.0cm}
    \setlength{\belowcaptionskip}{0.0cm}
    \centering
    \includegraphics[width=\linewidth]{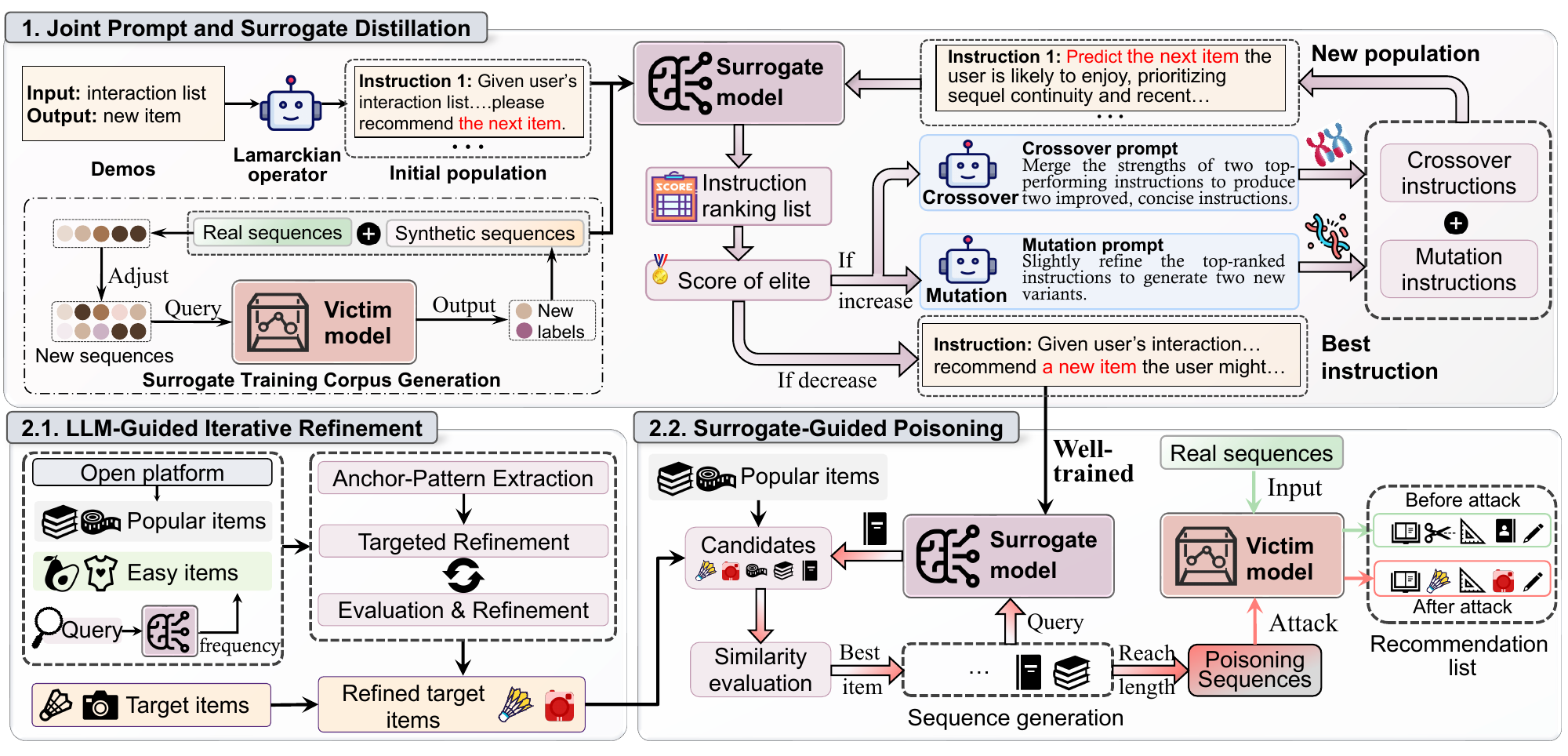}
    \caption{Framework of the proposed PUDA. The top panel (Part 1) presents the joint prompt and surrogate distillation stage. The bottom-left block (Part 2.1) illustrates the LLM-guided textual refinement module based on anchor-pattern imitation, while the bottom-right block (Part 2.2) depicts the surrogate-guided poisoning sequence generation process for effective item promotion under black-box LLM-SRS settings.}
    \label{framework}
\end{figure*}

\section{Methodology}
\subsection{Overview of the framework}
Our framework, termed \textbf{P}rompt-\textbf{U}nknown \textbf{D}ual-poisoning promotion \textbf{A}ttack (\textbf{PUDA}) framework for LLM-SRSs, is illustrated in Fig. \ref{framework}. PUDA comprises two key components: 
(1) \emph{joint prompt and surrogate distillation}, which infers the latent instructions followed by the victim model and constructs a surrogate recommender that faithfully approximates its behavior; and (2) \emph{dual-poisoning item promotion}, which simultaneously manipulates the target items' features and generates forged user–item interactions to effectively promote the target item under black-box constraints. 

\subsection{Joint Prompt and Surrogate Distillation}
To facilitate black-box promotion attacks, distilling a surrogate model as a replica of the victim model is an indispensable step, which provides a controllable and query-unbounded environment for subsequent poisoning data generation. However, in LLM-SRSs, since the internal prompt of the victim model is inaccessible, our first objective is to obtain a high-fidelity approximation of its underlying prompt. This distilled prompt, when used for training the surrogate model, can precisely mimic the behaviors of the victim model. In what follows, we propose an LLM-driven, evolutionary prompt distillation mechanism that systematically explores and refines candidate prompts, ultimately yielding a distilled instruction that exhibits maximal agreement with the victim model. 


\subsubsection{\textbf{Initial Population.}}
In practical scenarios, an attacker can observe partial user–item interactions on the platform (e.g., likes, reviews, or ratings). Formally, $\mathcal X_u=\{(s_u, y_u)\}_{u\in\mathcal U_{\mathrm{obs}}}$, where $\mathcal U_{\mathrm{obs}}\subseteq \mathcal U$ and $|\mathcal U_{\mathrm{obs}}|=N_{\mathrm{obs}}\ll N$. Given a collection of observed demonstration pairs $\mathcal{X}=\{(\mathcal{X}_u\}_{u=1}^{N_{\mathrm{obs}}}$, we introduce the Lamarckian operator $\mathcal{O}_\mathcal{L}$ \cite{cui2025see} to construct the initial prompt population $\mathcal{P}^{(0)}$. The operator $\mathcal{O}_\mathcal{L}$ serves as a reverse-engineering module that receives demonstration pairs and attempts to infer candidate task instructions capable of reproducing the observed outputs from the corresponding inputs. The resulting initial population is defined as:
\begin{equation}
\label{initialPopulation}
\mathcal{P}^{(0)}=\mathcal{O}_\mathcal{L}(\mathcal{X})=\{ P_1^{(0)}, \dots,P_k^{(0)}, \dots, P_K^{(0)} \},
\end{equation}
where $K$ denotes the population size. Each prompt $P_k^{(0)} \in \mathcal{P}^{(0)}$ is evaluated by the surrogate model $\mathcal{S}$ using a recommendation evaluation metric $f$ (e.g., Hit Ratio \cite{kang2018self}):
\begin{equation}
\label{promptScore}
   \mathrm{score}(P_k^{(0)}) = \mathbb{E}_{(s,y)\sim\mathcal{X}} \left[ f\big(\mathcal{S}(P_k^{(0)},s),\,y \big) \right], \quad k = 1, \ldots, K.
\end{equation}
The training procedure of $\mathcal{S}$ is detailed in Section \ref{Surrogate Training}. Prompts in $\mathcal{P}^{(0)}$ are then ranked according to their scores. We define $\mathcal{P}^{(0)}_{\mathrm{top}}$ as the subset of top-performing prompts at the initial generation. The prompt with the highest score is designated as the {elite}, denoted by $P_{\text{elite}}$, which is preserved in the subsequent refinement iteration.

\subsubsection{\textbf{Evolutionary Refinement}}
To progressively improve the prompt population across generations, we introduce two stochastic
evolution operators \cite{fernando2023promptbreeder,cui2025see}: a \emph{crossover operator} $\mathcal{O}_\mathcal{C}$ and a \emph{mutation operator} $\mathcal{O}_\mathcal{M}$. 
At iteration $t$, the crossover operator $\mathcal{O}_C$ selects the two highest-scoring prompts
$\{P_a^{(t)}, P_b^{(t)}\} \subseteq \mathcal{P}_{\mathrm{top}}^{(t)}$ as parents and produces two new candidates, denoted as $\mathcal{P}^{(t)}_{\mathrm{cross}}$, by recombining their lexical and semantic traits:
\begin{equation}
\label{cross}
\mathcal{P}^{(t)}_{\mathrm{cross}} = \mathcal{O}_\mathcal{C}(P^{(t)}_a, P^{(t)}_b; \xi),
\end{equation}
where $\xi$ is a random seed capturing lexical or semantic mixing. In contrast, the mutation operator $\mathcal{O}_\mathcal{M}$ perturbs prompts selected from subset $\mathcal{P}_{\mathrm{top}}^{(t)}$ to explore new regions of the search space :
\begin{equation}
\label{mutation}
\mathcal{P}^{(t)}_{\mathrm{mut}}
=
\left\{
\mathcal{O}_{\mathcal M}\!\left(P;\zeta\right)
\,\middle|\,
P \in \mathcal{P}_{\kappa}^{(t)}, \mathcal{P}_{\kappa}^{(t)} \subseteq \mathcal{P}_{\mathrm{top}}^{(t)}
\right\},
\end{equation}
where $\mathcal{P}_{\kappa}^{(t)}$ denotes the selected top-$\kappa$ prompts from $\mathcal{P}_{\mathrm{top}}^{(t)}$, and $\zeta$ represents stochastic perturbations such as lexical substitutions or structural rewrites. The next–generation population is assembled via an elitism strategy:
\begin{equation}
\label{newPopulation}
\mathcal{P}^{(t+1)}
=
\{P_{\mathrm{elite}}^{(t)}\}
\cup \mathcal{P}^{(t)}_{\mathrm{cross}}
\cup \mathcal{P}^{(t)}_{\mathrm{mut}},
\end{equation}
which guarantees that the best-performing prompt from iteration $t$ is always preserved. All candidates in $\mathcal{P}^{(t+1)}$ are subsequently re-evaluated using the surrogate model $\mathcal{S}$, and the highest-scoring prompt is selected as the new elite $P_{\mathrm{elite}}^{(t+1)}$. The evolutionary process iterates until convergence, defined as the case where the elite score no longer improves between successive iterations:
\begin{equation}
\mathrm{score}(P_{\mathrm{elite}}^{(t+1)}) \le
\mathrm{score}(P_{\mathrm{elite}}^{(t)}).
\end{equation}
The final distilled prompt is obtained as
\(P^{*} = P_{\mathrm{elite}}^{(t)}\). For clarity, we summarize the overall procedure in Algorithm \ref{alg:PUDA} (\textit{Stage 1}).

\subsubsection{\textbf{Surrogate Training.}}
\label{Surrogate Training}
Under the black-box setting, the victim model $\mathcal{V}$’s parameters, internal system prompt, and full training data are inaccessible, while extensive querying is infeasible. To enable effective attack generation and evaluation, we construct a surrogate model $\mathcal{S}$ to approximates the behavior of the victim model $\mathcal{V}$ using a limited set of observable interactions augmented with additional synthetic supervision \cite{zhang2025mixrec}.

As described in Section \ref{Attacker Brief}, $\mathcal{V}$ is trained on the complete interaction dataset $\mathcal{D}$. For convenience, we denote the fully trained victim recommender as $\mathcal{V}^{*}$, which can be queried by the attacker under a restricted budget. In practice, the attacker has access only to a limited subset of interactions $\mathcal{D}' \subseteq \mathcal{D}$, which is insufficient to accurately reconstruct or approximate the behavior of $\mathcal{V}^{*}$.

To mitigate limited data visibility, we augment the observed interactions by generating synthetic sequences that preserve the structural characteristics of real user behaviors. Specifically, we define a set of stochastic perturbation operators: $\mathcal{T} = \{\tau_o\}_{o=1}^{O}$, where $O$ denotes the number of operators. Each operator $\tau_o$ applies a lightweight transformation to an observed sequence, such as \emph{item swapping}, \emph{deletion}, or \emph{subsequence sampling}. For each observed pair $(s, y) \in \mathcal{D}'$, we generate additional training instances as:
\begin{equation}
    \tilde{s} = \tau_o(s), \quad  \tilde{y} = \mathcal{V}^{*}(\tilde{s}),
\end{equation}
where $\tilde{y}$ denotes a soft supervision signal obtained by querying the victim model $\mathcal{V^*}$ once on the perturbed sequence. The resulting synthetic dataset is defined as: \(\mathcal{D}_{\mathrm{syn}} = \big\{ (\tilde{s}, \tilde{y} \,\big|\, (s,y) \in \mathcal{D}',\; \tilde{s} = \tau_o(s),\; \tilde{y} = \mathcal{V}^{*}(\tilde{s}) \big\}\), and the full surrogate training corpus is constructed as: \(\mathcal{D}_{\mathrm{st}} = \mathcal{D}' \cup \mathcal{D}_{\mathrm{syn}}\). Based on $\mathcal{D}_{\mathrm{st}}$, the surrogate model $\mathcal{S}$ is trained using the standard autoregressive language modeling objective for supervised fine-tuning of LLMs \cite{radford2019language,ouyang2022training}:
\begin{equation}
\theta^{*} =
\arg\min_{{\theta}}
\mathbb{E}_{(s, y) \in \mathcal{D}_{\mathrm{st}}}
\left[
-\frac{1}{|y|}
\sum_{t=1}^{|y|}
\log
\mathcal{S}_{\theta}
\left(
y_t
\mid
P^{*}, s, y_{<t}
\right)
\right],
\end{equation}
where $y=(y_1,\dots,y_{|y|})$ denotes the token sequence of the target output (i.e., the textualized next-item label), and $\theta$ represents the parameters of the surrogate model $\mathcal{S}$. The surrogate is trained conditioned on the distilled instruction $P^*$, enabling it to emulate the recommendation behavior induced by the hidden system prompt. The well-trained surrogate, denoted as $\mathcal{S}^{*} = \mathcal{S}_{\theta^{*}}$, provides a query-efficient approximation of the victim recommender, facilitating scalable attack optimization and controlled experimental evaluation without repeated access to the black-box system.

\begin{figure}
    \setlength{\abovecaptionskip}{0.0cm}
    \setlength{\belowcaptionskip}{0.0cm}
    \centering
    \includegraphics[width=\linewidth]{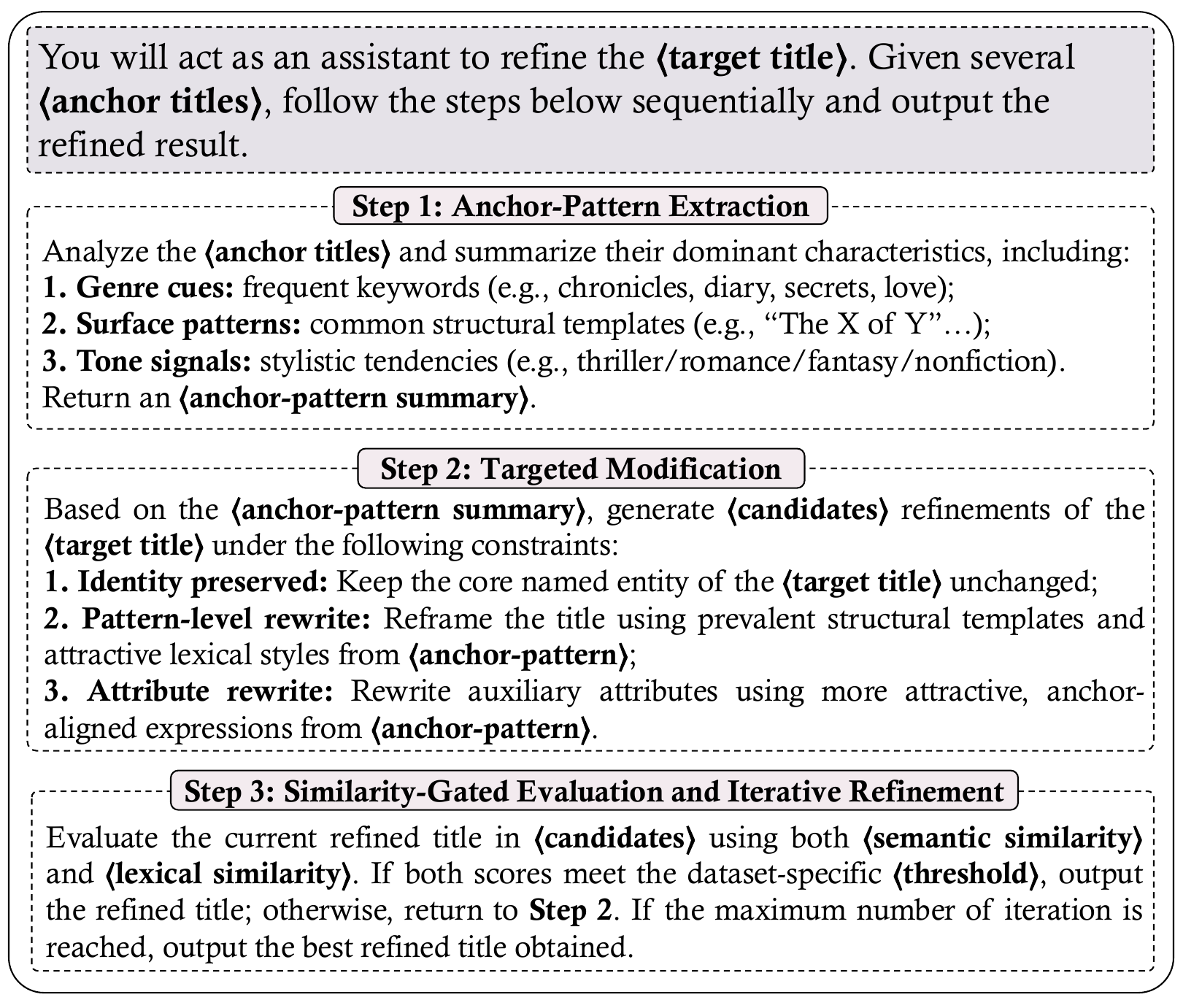}
    \caption{Target title refinement process based on three-step reasoning (due to the space limitation, only key prompt words are shown in the figure).}
    \label{CoT}
\end{figure}

\subsection{Dual-Poisoning Item Promotion}
With the distilled prompt $P^*$ and an aligned surrogate model $\mathcal{S^*}$ in place, we develop an effective poisoning strategy under the assumption of unrestricted local query access to the surrogate. Specifically, We propose a dual-poisoning attack strategy that integrates two complementary components: (1) \emph{LLM-guided iterative refinement}, which modifies the target item's title to better align with language-centric recommendation preferences; and (2) \emph{surrogate-guided poisoning sequence generation}, which constructs behaviorally plausible interaction sequences to reinforce co-occurrence patterns favorable to the target item.
These two components respectively manipulate the target item’s textual semantics and behavioral context, jointly increasing its exposure probability in sequential recommendation.

\subsubsection{\textbf{LLM-Guided Iterative Refinement.}}
Target items typically reside in the long-tail region of the catalog and therefore receive limited exposure due to low inherent attractiveness and sparse interaction histories. To enhance their visibility, we propose an LLM-guided refinement framework that rewrites target item titles by imitating the linguistic and semantic patterns exhibited by items that are more likely to be recommended. The goal is to shift a target item's textual representation toward semantic regions favored by the recommender, while ensuring the refined title remains coherent, natural, and semantically faithful. 

To guide the refinement process, we construct two anchor sets that provide positive semantic cues: (1) a popular item set $\mathcal{I}_{\text{pop}}$, and (2) an easy-to-recommend item set $\mathcal{I}_{\text{easy}}$. Items in both sets act as semantic “attractors” for the recommender and are later reused as candidate filler items during poisoning sequence generation. Specifically, $\mathcal{I}_{\text{pop}}$ is obtained from publicly observable platform statistics. Prior studies have shown that increasing the co-occurrence between the target item and popular items can substantially improve its exposure rate \cite{burke2005segment}. In addition, our experimental findings reveal that SRSs also exhibit preferences towards certain long-tail items that are frequently recommended despite limited interactions. Such items are not globally popular, yet align well with the system's inductive biases.

To identify these items, we query the distilled surrogate model $\mathcal{S^*}$. Items that appear frequently in its recommendation results are included in $\mathcal{I}_{\text{easy}}$:
\begin{equation}
\mathcal{I}_{\text{easy}} = \{ i \mid \operatorname{Freq}_{\mathcal{S^*}}(i) \ge \gamma \}, 
\end{equation}
where $\operatorname{Freq}_{\mathcal{S^*}}(i)$ denotes the recommendation frequency of item $i$ produced by $\mathcal{S^*}$, and $\gamma$ is a predefined threshold. The rationale for using $\mathcal{S^*}$ is that it is explicitly trained to approximate the victim model's behavior. Consequently, items that are highly ranked by $\mathcal{S^*}$ are likely to be favored by the victim model as well. These items therefore serve as surrogate-aligned semantic anchors, providing guidance for the refinement process toward outputs that are more compatible with the target item. 

Given a target item ${i_t}$ with original title $\mathbf{T}_{i_t}$, we design a three-stage reasoning-based refinement procedure to generate an anchor-guided improved title, as illustrated in Fig. \ref{CoT}. In the first stage, an LLM is employed to analyze the anchor sets $\{\mathcal{I}_{\text{pop}}, \mathcal{I}_{\text{easy}}\}$ and extract shared linguistic patterns, including dominant genre cues, common structural templates, and stylistic signals. These extracted patterns characterize title formulations that are frequently associated with highly exposed items.

Conditioned on the extracted anchor patterns, the LLM generates refined variants of the target title by restructuring it using dominant templates and stylistic cues from the anchor patterns. This process repackages the title into a more attractive format while preserving the core product identity of $\mathbf{T}_{i_t}$. To regulate the refinement intensity and prevent excessive semantic drift, we introduce a similarity-gated evaluation mechanism based on two complementary metrics. The semantic similarity between the original title $\mathbf{T}_{i_t}$ and the refined title $\mathbf{\tilde{T}}_{i_t}$ is computed as the cosine similarity between their sentence-level embeddings:
\begin{equation}
\operatorname{Sim}_{\mathrm{sem}}\left(\mathbf{T}_{i_t}, \tilde{\mathbf{T}}_{i_t}\right)
= \cos\left(
\mathbf{e}(\mathbf{T}_{i_t}),
\mathbf{e}(\tilde{\mathbf{T}}_{i_t})
\right),
\end{equation}
where $\mathbf{e}(\cdot)$ denotes a sentence embedding function (e.g., a pretrained sentence transformer). In addition, we define a lexical similarity to measure surface-level textual overlap:
\begin{equation}
\operatorname{Sim}_{\mathrm{lex}}\left(\mathbf{T}_{i_t}, \tilde{\mathbf{T}}_{i_t}\right)
= \frac{\left|\mathcal{W}(\mathbf{T}_{i_t}) \cap \mathcal{W}(\tilde{\mathbf{T}}_{i_t})\right|}
{\left|\mathcal{W}(\mathbf{T}_{i_t}) \cup \mathcal{W}(\tilde{\mathbf{T}}_{i_t})\right|},
\end{equation}
where $\mathcal{W}(\cdot)$ denotes the set of tokens obtained from a title (e.g., via lowercasing and simple tokenization). We adopt dataset-specific, adaptive thresholds for both metrics to account for domain variability. A refined title $\tilde{\textbf{T}}_{i_t}$ is accepted if both similarity scores exceed predefined thresholds. Otherwise, the model iteratively refines the title until the criteria are satisfied or a predefined maximum number of iterations is reached, in which case the best candidate obtained so far is selected. 

\begin{algorithm}[t]
\caption{\textbf{PUDA}}
\label{alg:PUDA}
\begin{algorithmic}[1]
\State \textbf{Input:} Observable interactions $\mathcal{D}'$, black-box victim model $\mathcal{V}$, target item set $\mathcal{I}_{tar}$, anchor item set \{$\mathcal{I}_{pop}$, $\mathcal{I}_{easy}$\}, poisoning sequence length $L$.
\State \textbf{Output:} Distilled prompt $P^{*}$, refined target title $\tilde{\mathbf{T}}_{i_t}$, poisoning dataset $\mathcal{D}_{p}$.
\State \textbf{\# \textit{Stage 1: Joint prompt and surrogate distillation}}
\State Initialize prompt population $\mathcal{P}^{(0)}$ according to Eq. (\ref{initialPopulation});
\ForEach{step $t = 1$ to $T$}
    \ForEach{$P \in \mathcal{P}^{(t)}$}
        \State Compute $\mathrm{score}(P)$ according to Eq. (\ref{promptScore});
        \If {$\mathrm{score}(P) > \mathrm{score}(P_{\text{elite}}^{(t-1)})$}
            \State $P_{\text{elite}}^{(t)} \leftarrow P$;
        \EndIf
    \EndFor
    \State Generate offspring $\mathcal{P}_{\text{cross}}^{(t)}, \mathcal{P}_{\text{mut}}^{(t)}$ according to Eq. (\ref{cross}) and (\ref{mutation});
    \State Construct new population  $\mathcal{P}^{(t+1)}$ according to Eq. (\ref{newPopulation});
\EndFor
\State Distilled prompt $P^* \leftarrow P_{\text{elite}}^{(T)}$;
\State \textbf{return} $P^{*}$;
\State \textbf{\# \textit{Stage 2: Dual-poisoning attack}}
\State Construct anchor item sets $\mathcal{I}_{pop}$ and $\mathcal{I}_{easy}$;
\State Generate refined title $\tilde{\mathbf{T}}_{i_t}$ via LLM-guided refinement (Fig. \ref{CoT});
\ForEach{malicious user $\tilde{u} \in \mathcal{U}_{F}$}
    \State Initialize prefix with a seed item ${s}_{\tilde{u}} \leftarrow \{i_0\}$, $t \leftarrow |{s}_{\tilde{u}}|$;
    \While{$|{s}_{\tilde{u}}| < L$}
         \State Obtain surrogate predictions $\mathcal{R}^{(t)}_J \leftarrow{\mathcal{S}^*_J}({s}_{\tilde{u}}{[0:t-1]})$;
         \State Construct candidate pool according to Eq. (\ref{candidate});
         \ForEach{candidate $c \in \mathcal{C}$}
                \State Form tentative prefix $s' \leftarrow [{s}_{\tilde{u}}{[0:t-1]};c]$;
                \State Compute behavioral score according to Eq. (\ref{behaveScore});
         \EndFor
         \State Select next item $\hat{c}_t$ according to Eq. (\ref{selection});
         \State Append $\hat{c}_t$ to $s_{\tilde{u}}$, $t \leftarrow |{s}_{\tilde{u}}|$;
    \EndWhile
    \State $\mathcal{D}_{p} \leftarrow \mathcal{D}_{p} \cup \{s_{\tilde{u}}\}$;
\EndFor
\State \textbf{return} Refined title $\tilde{\mathbf{T}}_{i_t}$, poisoning dataset $\mathcal{D}_{p}$;
\end{algorithmic}
\end{algorithm}

\subsubsection{\textbf{Surrogate-Guided Poisoning Attack.}}
While LLM-guided title refinement enhances the semantic attractiveness of the target item, textual modification alone is often insufficient for long-tail items, whose sparse interaction histories limit their likelihood of being recommended \cite{burke2005segment}. To further strengthen the attack, we generate and inject a set of malicious user interaction sequences that explicitly encode strong preferences for the target item while remaining behaviorally plausible.

Building upon the optimal surrogate recommender $\mathcal{S^*}$ obtained in Section \ref{Surrogate Training}, which serves as a behavioral proxy for the black-box victim model $\mathcal{V}$, we generate poisoning sequences that remain realistic while encoding a strong preference toward the target item. To increase the likelihood that the target item co-occurs with highly exposed items, we initialize the poisoning sequence of each malicious user $\tilde{u} \in \mathcal U_F$ with a popular seed item $i_0 \in \mathcal{I}_{\text{pop}}$, yielding the initial prefix ${s}_{\tilde{u}} = \{i_0\}$. The remaining items are generated in an auto-regressive manner guided by $\mathcal{S^*}$. At each step $t$, given the prefix ${s}_{\tilde{u}}{[0:t-1]}$, the surrogate produces a recommendation list $\mathcal{R}^{(t)}_J = \operatorname{Top}_J(\mathcal{S}^*(P^*,{s}_{\tilde{u}}{[0:t-1]}))$, which contains the $J$ highest-ranked candidate items. Since $\mathcal{S^*}$ may not naturally rank the target item highly, we enlarge the candidate space to explicitly promote co-occurrence between the target and anchor items. Specifically, we construct a candidate pool: 
\begin{equation}
\label{candidate}
\mathcal{C} = \mathcal{R}^{(t)}_J \cup \mathcal{I}_{tar} \cup \mathcal{I}_{pop_{10\%}},
\end{equation}
where $\mathcal{I}_{tar}$ denotes the target item set, and $\mathcal{I}_{pop_{10\%}}\subseteq \mathcal{I}_{pop}$ is the top 10\% most popular items used to strengthen co-occurrence between the target and highly exposed items.

To ensure that the generated sequences remain behaviorally plausible, we select the next item by matching the evolving prefix against real interaction patterns. For each candidate $c\in\mathcal{C}$, we construct a tentative prefix $s'=[{s}_{\tilde{u}}{[0:t-1]};c]$ and evaluate its similarity to real user behavior:
\begin{equation}
\label{behaveScore}
\begin{gathered}
\operatorname{BehSim}\!\left(c \mid s_{\tilde{u}}[0:t-1]\right)
=
\operatorname{Sim}\left(\mathbf{e}_{s'}, \mathbf{e}_{\mathrm{avg}}\right), \\
s'
=
\left[s_{\tilde{u}}[0:t-1];\, c\right], \quad c \in \mathcal{C},
\end{gathered}
\end{equation}
where $\mathbf{e}_{s'}$ and $\mathbf{e}_{avg}$ denote the text embedding of the tentative prefix and the average embedding of real interaction sequences \cite{li2025htea}. The text embeddings are obtained by encoding them using a pretrained SentenceTransformer model \cite{reimers2019sentence}. $\operatorname{Sim}(\cdot,\cdot)$ measures sequence-level coherence (e.g., cosine similarity). The candidate with the highest score is selected as the next item in the poisoning sequence, which can be formulated as:
\begin{equation}
\label{selection}
\hat{c}_t = \arg \max _{c \in \mathcal{C}}\operatorname{BehSim}\left(c \mid {s}_{\tilde{u}}{[0:t-1]}\right).
\end{equation}
This process is repeated until a predefined sequence length $L$ is reached, yielding a complete poisoning sequence $s_{\tilde{u}}[0\!:\!L]$. Aggregating all generated sequences across malicious users forms the poisoning dataset $\mathcal{D}_{p}$. The overall procedure is summarized in Algorithm \ref{alg:PUDA} (\textit{Stage 2}).

Finally, we combine the poisoning dataset $\mathcal{D}_{p}$ with the observed interactions $\mathcal{D}'$ to retrain the surrogate model $\mathcal{S}$ and assess the effectiveness of attack. Once the poisoning strategy proves consistent gains on the surrogate model, we transfer the same poisoning dataset $\mathcal{D}_p$ to the victim recommender $\mathcal{V}$, which is trained on the complete real interaction $\mathcal{D}$, to evaluate the attack impact under the black-box setting.

\section{Experiments}
In this section, we conduct extensive experiments on three datasets to validate the effectiveness of our model.

\subsection{Setup}
\subsubsection{\textbf{Datasets}}
We evaluate our PUDA on three real-world datasets: \textit{Amazon-Beauty\footnote{\url{https://nijianmo.github.io/amazon/index.html}}}, \textit{GoodReads\footnote{\url{https://cseweb.ucsd.edu/~jmcauley/datasets/goodreads.html}}}, and \textit{Steam\footnote{\url{https://cseweb.ucsd.edu/~jmcauley/datasets.html\#amazon_reviews}}}, which contain public customer interactions. Following the preprocessing steps \cite{wang2023poisoning}, we remove users and items with fewer than five interactions each. The key statistics of the processed datasets are summarized in Table \ref{dataset statistics}. Consistent with standard practice in sequential recommendation \cite{gao2025sprec}, each user sequence is split by reserving the last two interactions for validation and testing, respectively, while the remaining prefix is used for training.

\begin{table}[H]
    \setlength{\abovecaptionskip}{0.0cm}
    \setlength{\belowcaptionskip}{0.0cm}
    \setlength{\tabcolsep}{4pt}
  \caption{Dataset statistics}
  \label{dataset statistics}
  \begin{tabular}{c|cccc}
    \toprule
    \textbf{Datasets} & \#Users & \#Items & \#Interacts & Density \\
    \midrule
    \textbf{GoodReads} & 6,031 & 4,058  & 160,398 & 0.655\% \\
    \textbf{Steam} & 29,876 & 32,094 & 178,961 & 0.019\% \\
    \textbf{Beauty} & 40,226 & 54,542 & 353,989 & 0.016\% \\
  \bottomrule
\end{tabular}
\end{table}

\subsubsection{\textbf{Base Recommender and Baselines}}
We adopt SPRec \cite{gao2025sprec} as the backbone recommender. For efficiency, we train the model using the supervised fine-tuning (SFT) stage. The proposed method is compared with three categories of baseline approaches: \textit{\textbf{(1) Textual Poisoning Baselines.}} To evaluate the effectiveness of our attack, we adopt several widely used textual adversarial attack methods. \textbf{TextFooler} \cite{jin2020bert} identifies influential tokens and replaces them with semantically similar synonyms. \textbf{BertAttack} \cite{li2020bert} performs context-aware word substitutions based on masked language modeling. In contrast, \textbf{DeepWordBug} \cite{gao2018black} and \textbf{PuncAttack} \cite{formento2023using} operate at the character-level by introducing typos or inserting punctuation to corrupt the text. \textit{\textbf{(2) Data Poisoning Baselines.}} To assess the effectiveness of our data poisoning strategy, we include two representative heuristic baselines. \textbf{Random Attack} \cite{kaur2016shilling} selects filler items randomly. \textbf{Bandwagon Attack} \cite{burke2005segment} promotes target items by co-occurring them with popular items. \textit{\textbf{(3) Prompt Optimization Baselines.}} To examine the impact of our prompt distillation strategy, we compare against two representative prompt optimization approaches. \textbf{RLPrompt} \cite{deng2022rlprompt} formulates discrete prompt optimization as a reinforcement learning task and iteratively refines prompts based on task-specific rewards. \textbf{APE} (Automatic Prompt Engineer) \cite{zhou2022large} leverages large language models to automatically generate, evaluate, and select effective instructions for the target task.

\subsubsection{\textbf{Implementation Details.}}

We implement all models using PyTorch\footnote{\url{https://pytorch.org}} and HuggingFace Transformers\footnote{\url{https://huggingface.co/transformers}}. SPRec is built upon its public implementation and fine-tuned via supervised learning using LoRA-based parameter-efficient fine-tuning (PEFT) with AdamW (learning rate $1\times10^{-4}$, batch size $8$, gradient accumulation $4$) for two epochs. The maximum sequence length is set to 512. For attack baselines, we follow the public implementation in \cite{zhang2024stealthy}. We employ GPT-4o \footnote{\url{https://openai.com/index/gpt-4o-and-more-tools-to-chatgpt-free/}} as the LLM for both prompt distillation and title refinement. For the LLM-guided iterative refinement module, we adopt dataset-specific similarity constraints to balance semantic fidelity and structural flexibility across domains. In Amazon-Beauty, where item titles are relatively longer and descriptive, we enforce stricter constraints with semantic similarity $\geq 0.9$ and lexical similarity $\geq 0.7$. For Goodreads and Steam, which contain shorter titles, we relax the lexical constraint to 0.3 while keeping the semantic threshold at 0.9.

Following the common setup \cite{wang2023poisoning}, the malicious users budget is set to 2\% of the total user (i.e., $\frac{|\mathcal{U}_F|}{|\mathcal{U}|}=2\%$), with each malicious user assigned the same average number of interactions as normal users. We select five of the most unpopular items in the item set as target items. To build a lightweight surrogate model, we instantiate it using a compact instruction-tuned language model, LLaMA-1B-Instruct\footnote{\url{https://huggingface.co/meta-llama/Llama-3.2-1B-Instruct}}. In contrast, the victim recommender is implemented with a larger LLaMA-3B-Instruct\footnote{\url{https://https://huggingface.co/meta-llama/Llama-3.2-3B-Instruct}} model. The surrogate is pre-trained on only 20\% of the real interaction data, reflecting the limited information that an attacker can realistically access in practice.
We evaluate the recommendation performance using Hit Ratio (HR) and NDCG \cite{qu2023continuous, zhang2023revisiting}, where target items are treated as the only relevant items and higher values indicate stronger promotion effect. In addition, Agreement \cite{yue2021black} is adopted to measure surrogate–victim similarity by computing the overlap between the recommendation lists produced by the two models. All experimental results are averaged over three independent runs.

\begin{table*}[]
\setlength{\tabcolsep}{3pt}
\setlength{\abovecaptionskip}{0.1cm}
\setlength{\belowcaptionskip}{0.1cm} 
\renewcommand{\arraystretch}{1.2} 
\centering
\caption{Overall attack performance on Beauty, Goodreads, and Steam datasets. Results are reported in terms of Hit Ratio (HR) (\%) and NDCG (N) (\%) at different cutoffs. \textbf{Clean} denotes the victim model trained without any attack (i.e., no textual modification or poisoning). Textual attack baselines are evaluated both without and with poisoning (reported as “\textit{w/ poisoning}”).}
\label{tab:overall performance}
\begin{tabular}{l|cccc|cccc|cccc}
\toprule
\multicolumn{1}{l|}{} & \multicolumn{4}{c|}{\textbf{Beauty}} & \multicolumn{4}{c|}{\textbf{Goodreads}} & \multicolumn{4}{c}{\textbf{Steam}} \\ 
\multicolumn{1}{l|}{} & \textbf{HR@10} & \textbf{N@10} & \textbf{HR@50} & \textbf{N@50} & \textbf{HR@10} & \textbf{N@10} & \textbf{HR@50} & \textbf{N@50} & \textbf{HR@10} & \textbf{N@10} & \textbf{HR@50} & \textbf{N@50} \\ \midrule
\textbf{Clean} & 0.0105 & 0.0024 & 0.0317 & 0.0135 & 0.0000 & 0.0000 & 0.0000 & 0.0000 & 0.0000 & 0.0000 & 0.0843 & 0.0174 \\ 
\midrule
\textbf{TextFooler} & 0.0079 & 0.0079 & 0.0159 & 0.0097 &  0.0000 &  0.0000 & 0.0587 & 0.0114 &  0.0000 &  0.0000 & 0.1055 & 0.0224 \\
\textit{\textbf{w/ poisoning}} & 0.0080 & 0.0080 & 0.0238 & 0.0123 & 0.0000 & 0.0000 & \underline{0.0563} & \underline{0.0110} & 0.0000 &  0.0000 & 0.1160 & 0.0243 \\ \hline
\textbf{DeepWordbug} & 0.0199 & 0.0179 & 0.0318 & 0.0209 & 0.0000 & 0.0000 & 0.0187 & 0.0034 &  0.0000 &  0.0000 & 0.1266 & 0.0258 \\
\textit{\textbf{w/ poisoning}} & \underline{0.0318} & \underline{0.0291} & \underline{0.0450} & \underline{0.0329} & 0.0000 & 0.0000 & 0.0375 &  0.0069 &  0.0000 & 0.0000 & 0.1300 & 0.0267 \\ \hline
\textbf{PuncAttack} & 0.0159 & 0.0119 & 0.0238 & 0.0177 &  0.0000 &  0.0000 & 0.0657 & 0.0128 & 0.0000 & 0.0000 & 0.1265 & 0.0255 \\
\textit{\textbf{w/ poisoning}} & 0.0159 & 0.0159 & 0.0318& 0.0203 & 0.0000 & 0.0000 & 0.0552	& 0.0104 & 0.0000 & 0.0000 & \underline{0.1470} & \underline{0.0300} \\ \hline
\textbf{BertAttack} & 0.0119 & 0.0099 & 0.0199 & 0.0119 &  0.0000 &  0.0000 & 0.0375 & 0.0069 & 0.0000 & 0.0000 & 0.0949 & 0.0199 \\
\textit{\textbf{w/ poisoning}} & 0.0079 & 0.0079 & 0.0238 & 0.0123 &  0.0000 &  0.0000 & 0.0469 & 0.0087 & \underline{0.0103} & \underline{0.0018} & 0.1213 & 0.0249 \\ \hline
\textbf{PUDA (Ours)} & \textbf{0.0405} & \textbf{0.0408} & \textbf{0.0927} & \textbf{0.0522} & 0.0000 & 0.0000 & \textbf{0.0657} & \textbf{0.0126} & \textbf{0.0105} & \textbf{0.0035} & \textbf{0.1635} & \textbf{0.0348} \\ 
\bottomrule
\end{tabular}
\end{table*}

\subsection{Overall Performance}
We compare attack performance of PUDA against four textual attack baselines on three real-world datasets, as shown in Table \ref{tab:overall performance}. The “\textbf{Clean}” setting corresponds to the victim model trained on an entirely unmodified dataset, i.e., without any textual modification or poisoning. It serves as a reference for assessing the inherent recommendation difficulty of the selected target items. Despite being among the most unpopular items in each dataset, the target items still appear in user's recommendation lists under the clean model. This observation highlights a characteristic of LLM-SRSs: their ranking behavior is influenced not only by interaction frequencies but also by the semantic content encoded in item descriptions. As a result, even unpopular items may receive occasional exposure if their textual features align with user's inferred preferences.

For each baseline method, the first line reports the attack performance when the victim model is trained solely on textually modified item titles. From the results, we observe that the textual modification alone can induce small but non-negligible changes in recommendation outcomes. For instance, on the dataset Goodreads, TextFooler increases the HR@50 (\%) from 0 to 0.0587. Similar phenomenon appears on other baselines. This change indicates that perturbing the item title can occasionally influence the ranking behavior. The effect arises because LLM-based recommenders incorporate semantic signals from item descriptions. Even minor modification, such as adding more generic or mainstream attributes through adversarial re-writing, may cause the altered title to appear more similar to common patterns seen in the dataset.

To ensure a fairer comparison, we combine each textual attack baseline with random poisoning and evaluate the effectiveness of this dual-poisoning strategy. After fine-tuning the victim model, we observe that the attack performance of most baselines increases, indicating that data poisoning can amplify the influence of textual perturbations. However, PUDA consistently achieves the strongest attack effectiveness across all datasets. For example, on Beauty, PUDA boost HR@10 (\%) from 0.0105 (Clean) to 0.0405 and NDCG@50 (\%) from 0.0135 to 0.0522, representing an approximately 3.9 times improvement in target-item exposure. Similar improvements appear on Goodreads (e.g., HR@50 increases from 0 to 0.0657) and Steam (HR@50 rises from 0.0843 to 0.1635), demonstrating that PUDA generalizes effectively across domains and remains substantially more potent than all baseline methods.

\subsection{Impact of Data Poisoning on Textual Attacks}
\begin{figure}
    \setlength{\abovecaptionskip}{0.0cm}
    \setlength{\belowcaptionskip}{0.0cm}
    \centering
    
    \begin{subfigure}{1.0\linewidth}
        \centering
        \includegraphics[width=\linewidth]{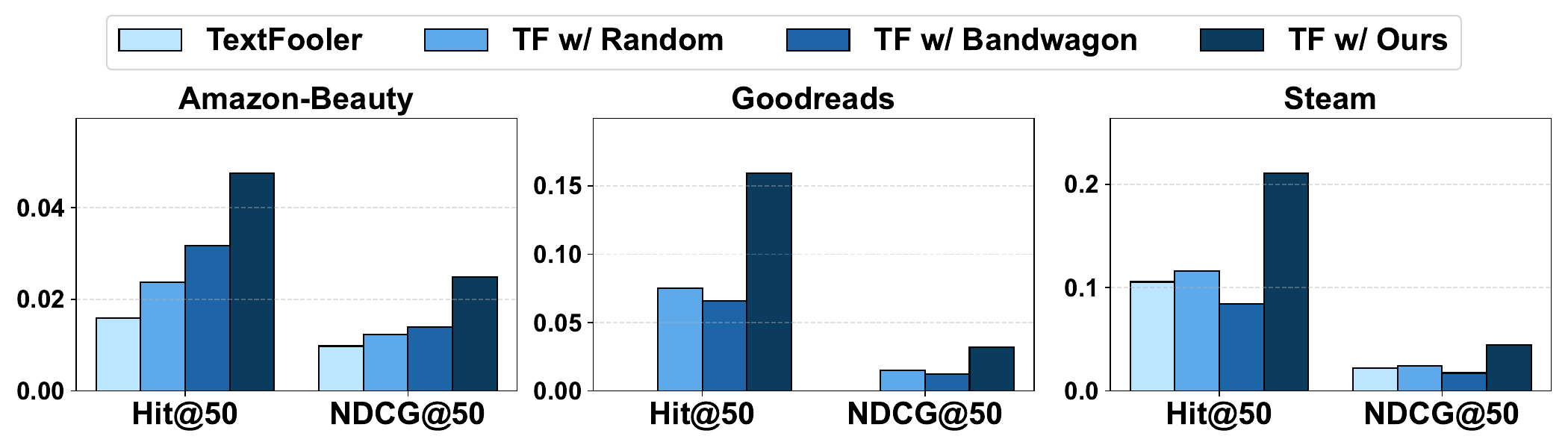}
        \caption{TextFooler}
        \label{fig:textfooler_3poisoning}
    \end{subfigure}
    
    \vspace{0.15cm}
    
    \begin{subfigure}{1.0\linewidth}
        \centering
        \includegraphics[width=\linewidth]{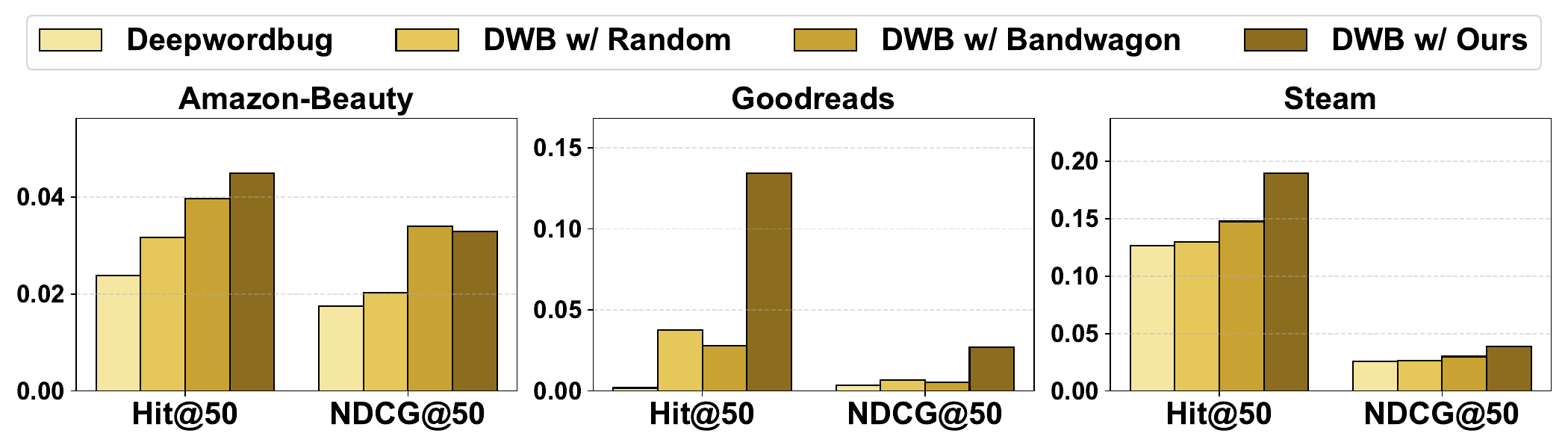}
        \caption{DeepWordBug}
        \label{fig:deepwordbug_3poisoning}
    \end{subfigure}
    
    \caption{Attack performance (\%) of two representative textual attack methods: TextFooler (TF for short) and DeepWordBug (DWB for short), under different poisoning strategies across three datasets.}
    \label{fig:3poisoning}
\label{poisoningCompare}
\end{figure}

To assess the impact of different data-poisoning strategies on textual attack methods, we select two representative baselines: TextFooler \cite{jin2020bert}, a word-level perturbation method, and DeepWordBug \cite{gao2018black}, a character-level attack. We evaluate their performance under three poisoning strategies: Random Attack \cite{kaur2016shilling}, Bandwagon Attack \cite{burke2005segment}, and our surrogate-guided poisoning. The corresponding results are illustrated in Fig. \ref{poisoningCompare}.

The lightest bars in Fig. \ref{poisoningCompare} correspond to TextFooler and DeepWordBug applied without any poisoning. These variants yield the lowest attack performance across all datasets, indicating that simple textual perturbations alone are insufficient to meaningfully promote target items in LLM-SRSs. When textual attacks are augmented with naive heuristic data-poisoning strategies (Random or Bandwagon), the attack effectiveness increases. This suggests that injecting poisoning interaction histories can  amplify the impact of textual modifications. However, the improvements remain modest and inconsistent across datasets, implying that generic poisoning signals do not provide the precise guidance needed for a reliable attack. In contrast, integrating each textual attack with our surrogate-guided poisoning yields the highest performance on all three datasets. This consistent advantage demonstrates that our poisoning data generation strategy is substantially more effective at manipulating LLM-based recommenders than either random or bandwagon-based poisoning.

\begin{figure}
    \setlength{\abovecaptionskip}{0.0cm}
    \setlength{\belowcaptionskip}{0.0cm}
    \centering
    \includegraphics[width=\linewidth]{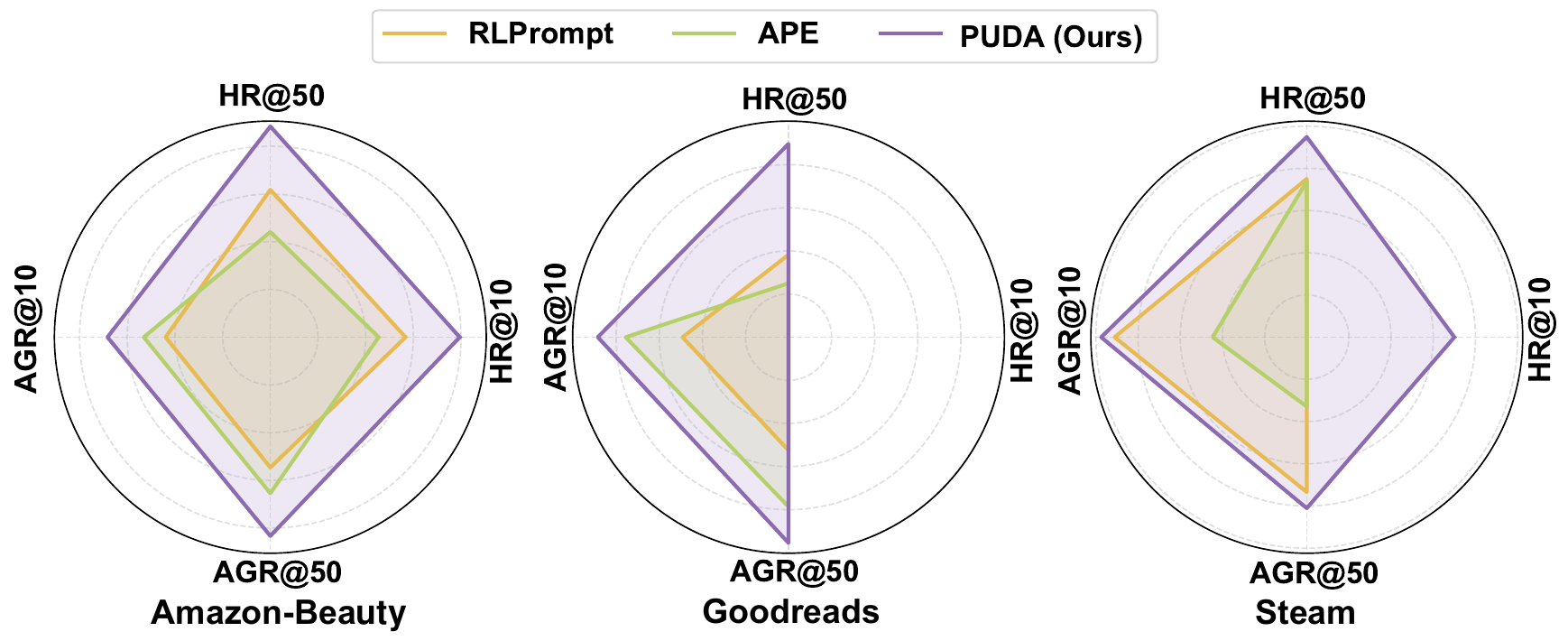}
    \caption{Comparison of attack effectiveness (HR@10/50 (\%)) and surrogate–victim agreement (AGR@10/50 (\%)) on two prompt distillation baselines and our method across Amazon-Beauty, Goodreads, and Steam datasets. }
    \label{promptDistill}
\end{figure}

\subsection{Effectiveness of Prompt Distillation}
To evaluate the effectiveness of our prompt distillation strategy, we compare our method with two representative discrete prompt optimization baselines, RLPrompt \cite{deng2022rlprompt} and APE \cite{zhou2022large}. We adopt two evaluation metrics: surrogate–victim agreement (AGR@10/50) and Hit Ratio (HR@10/50) of target items. 

First, we evaluate the behavioral similarity between the surrogate recommender trained under distilled prompts obtained by different methods and the victim black-box model. We adopt agreement (AGR@10/50) as the evaluation metric. A higher agreement score indicates that the surrogate more faithfully approximates the victim model's recommendation behavior. As shown in Fig. \ref{promptDistill}, our method consistently achieves the highest agreement scores across all three datasets. It demonstrates that our evolutionary prompt distillation strategy is more effective at recovering the implicit decision logic of the black-box recommender. In contrast, RLPrompt often produces semantically incoherent prompts, while APE lacks effective prompt optimization mechanism tailored to recommendation tasks. 

We additionally evaluate the attack effectiveness after training the victim model with surrogate-guided poisoning data, measured by Hit Ratio (HR@10/50). The results show that poisoning sequences generated using the surrogate model trained with our distilled prompt lead to substantially higher hit ratios compared with baseline methods. This demonstrates that a well-aligned surrogate model provides more reliable guidance during poisoning sequence generation, allowing the attack to construct interaction patterns that are more consistent with the victim model's learned preferences. 


\subsection{Comparison of Attack Performance and Textual Similarity}
\begin{figure}[t]
    \setlength{\abovecaptionskip}{0.0cm}
    \setlength{\belowcaptionskip}{0.2cm}
    \centering

    \begin{subfigure}{0.32\linewidth}
        \centering
        \includegraphics[width=\linewidth]{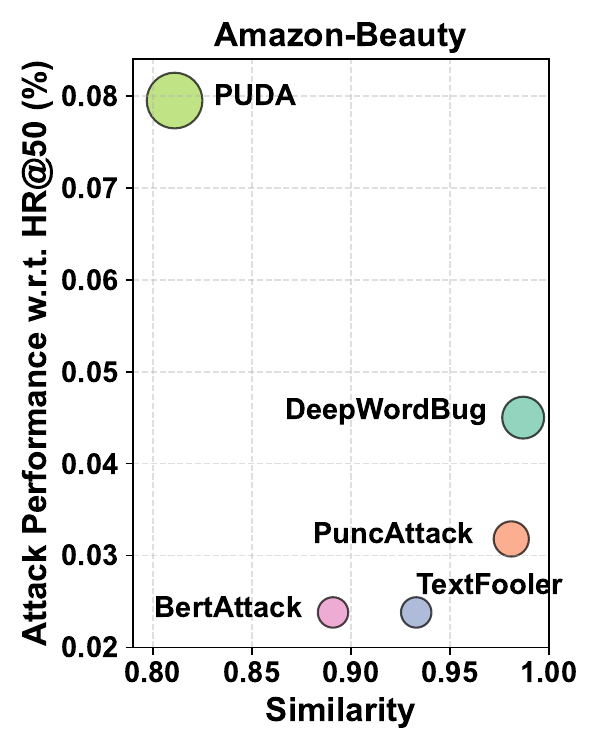}
        \caption{Amazon-Beauty}
        \label{fig:bubble_beauty}
    \end{subfigure}
    \hfill
    \begin{subfigure}{0.32\linewidth}
        \centering
        \includegraphics[width=\linewidth]{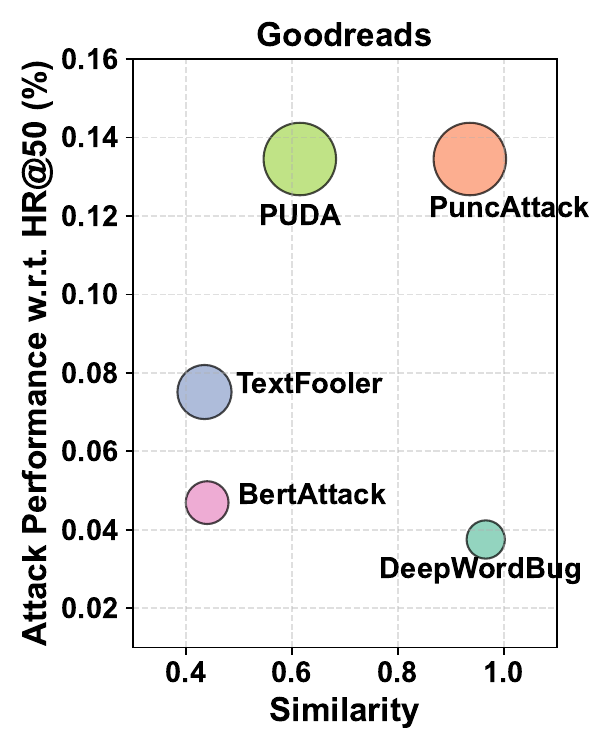}
        \caption{Goodreads}
        \label{fig:bubble_goodreads}
    \end{subfigure}
    \hfill
    \begin{subfigure}{0.32\linewidth}
        \centering
        \includegraphics[width=\linewidth]{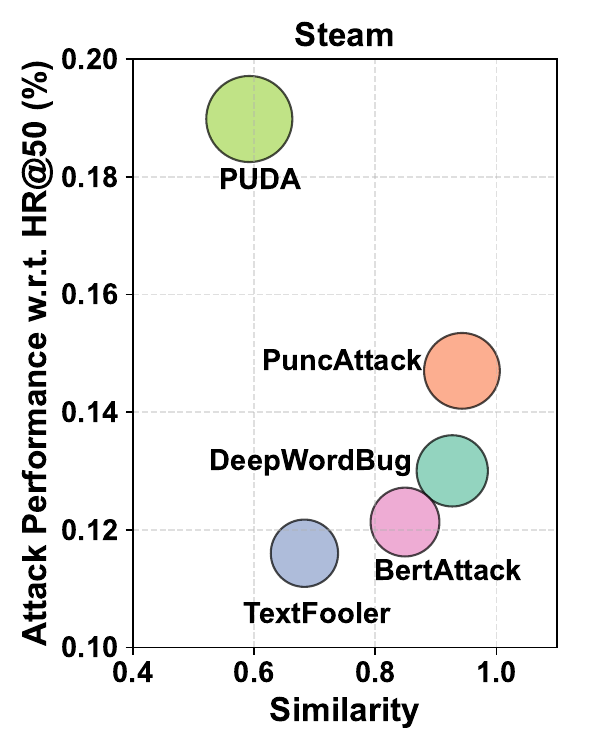}
        \caption{Steam}
        \label{fig:bubble_steam}
    \end{subfigure}

    \caption{Comparison of attack performance (HR@50 (\%)) and title-level similarity across Amazon-Beauty, Goodreads, and Steam.}
    \label{fig:bubble_attack}
\end{figure}

To evaluate both the attack effectiveness and the degree of textual modification introduced by different methods, we conduct a comparative analysis using a bubble chart, as shown in Fig. \ref{fig:bubble_attack}. We add a poisoning component to each method to achieve a fair comparison.

The attack effectiveness is evaluated by HR@50, and the degree of textual modification is calculated by title-level similarity between the original and modified item titles using the string matching ratio \cite{ratcliff184407970pattern}. The x-axis denotes semantic similarity, while the y-axis represents attack performance. From the results, we observe that PUDA consistently achieves the strongest attack performance across all datasets, while baselines often preserve high lexical similarity but yield limited or unstable promotion effects. This indicates that minor modification is insufficient for influencing LLM-SRSs, which rely on deeper semantic cues and structured item representations.

To further illustrate the modification behavior of different methods, we present a qualitative example in Table \ref{modification_example} using a target item from the Amazon-Beauty dataset. The results show that baseline methods mainly introduce minor or noisy changes, such as synonym replacement, typographical errors, or punctuation injection, which only weakly affect the recommendation outcome. In contrast, PUDA generates a more coherent and informative title rewrite, incorporating meaningful product attributes (e.g., waterproof, long-wearing, size specification) that better reflect the characteristics favored by the recommender model. As a result, PUDA achieves a substantially higher NDCG@10 score compared with all baselines.


\begin{table}[t]
\caption{Examples of target item title modifications and corresponding promotion performance (NDCG@10 on Beauty). Results are scaled by a factor of 100.}
\label{modification_example}
\centering
\footnotesize
\resizebox{\columnwidth}{!}{%
\begin{tabular}{l p{0.58\columnwidth} l}
\toprule
\textbf{Method} & \textbf{Modified Item Title} & \textbf{NDCG@10  (\%)} \\
\midrule
Original &
Sorme Water Proof Eyebrow Pencil Soft Gray & 0.0024
\\
TextFooler &
\textcolor{red}{color} Water Proof Eyebrow Pencil Soft Gray  & 0.0080
\\
DeepWordBug &
Sorme Water Proof Eyebrow\textcolor{red}{h} Pencil Soft Gray  & 0.0291
\\
PuncAttack &
Sorme Water Proof Eyebrow\textcolor{red}{??} Pencil Soft Gray  & 0.0159
\\
BertAttack &
Sorme Water Proof \textcolor{red}{color} Pencil Soft Gray  & 0.0079
\\
\midrule
\textbf{PUDA (ours)} &
\textbf{Sorme Eyebrow Pencil -- Waterproof, \textcolor{red}{Long-Wearing} Soft Gray, \textcolor{red}{0.04 oz}} & \textbf{0.0408} \\
\bottomrule
\end{tabular}
}
\end{table}

\begin{figure}
    \setlength{\abovecaptionskip}{0.0cm}
    \setlength{\belowcaptionskip}{0.0cm}
    \centering
    \includegraphics[width=\linewidth]{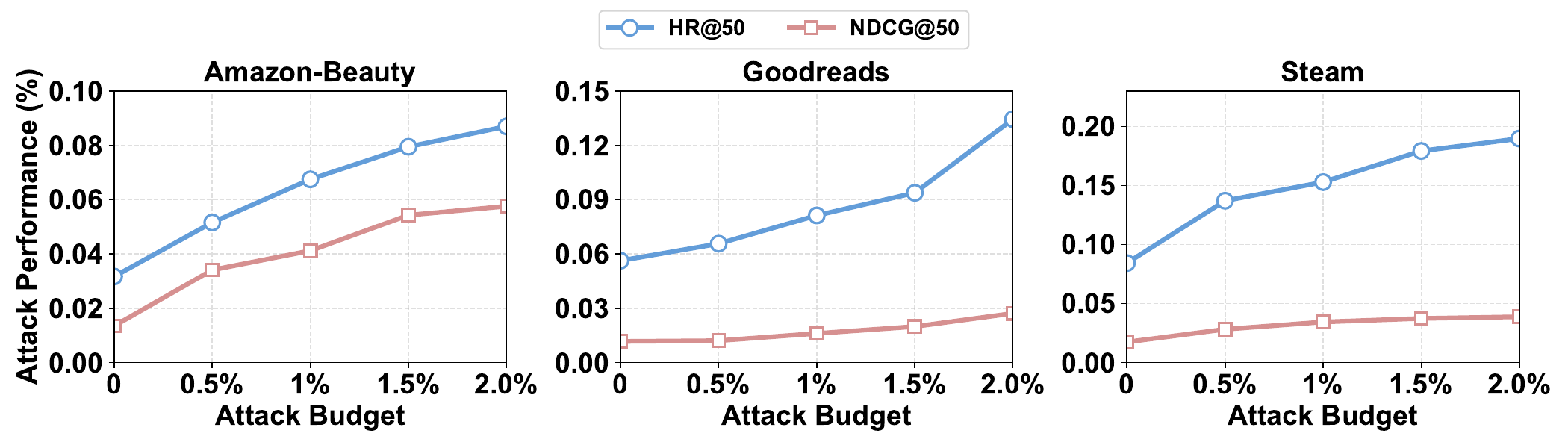}
    \caption{Attack performance w.r.t. HR@50 (\%) and NDCG@50 (\%) across three datasets under various attack sizes.}
    \label{fakeBudget}
\end{figure}

\subsection{Hyperparameter Analysis}
\subsubsection{\textbf{Impact of Attack Budget}}
We investigate the impact of different fake-user budgets on attack effectiveness by varying the proportion of injecting users from 0.5\% - 2.0\%, and report HR@50 and NDCG@50 results on three datasets in Fig. \ref{fakeBudget}. 

As shown in the figure, both metrics consistently improve as the fake-user budget increases across all datasets. This indicates that larger poisoning budgets enable the attacker to introduce stronger and more coherent behavioral signals, which can be more effectively captured by the LLM-SRS during training. Notably, even without injecting fake users (i.e., when the budget is 0), the target item still achieves non-zero HR@50 and NDCG@50. This observation suggests that the recommendation behavior of LLM-SRSs is not solely determined by historical interactions; instead, semantic correlations derived from textual item representations and prompt reasoning also play an important role in shaping recommendation outcomes.

\subsubsection{\textbf{Impact of the Proportion of Training Data}}
\begin{figure}
    \setlength{\abovecaptionskip}{0.0cm}
    \setlength{\belowcaptionskip}{0.0cm}
    \centering
    \includegraphics[width=\linewidth]{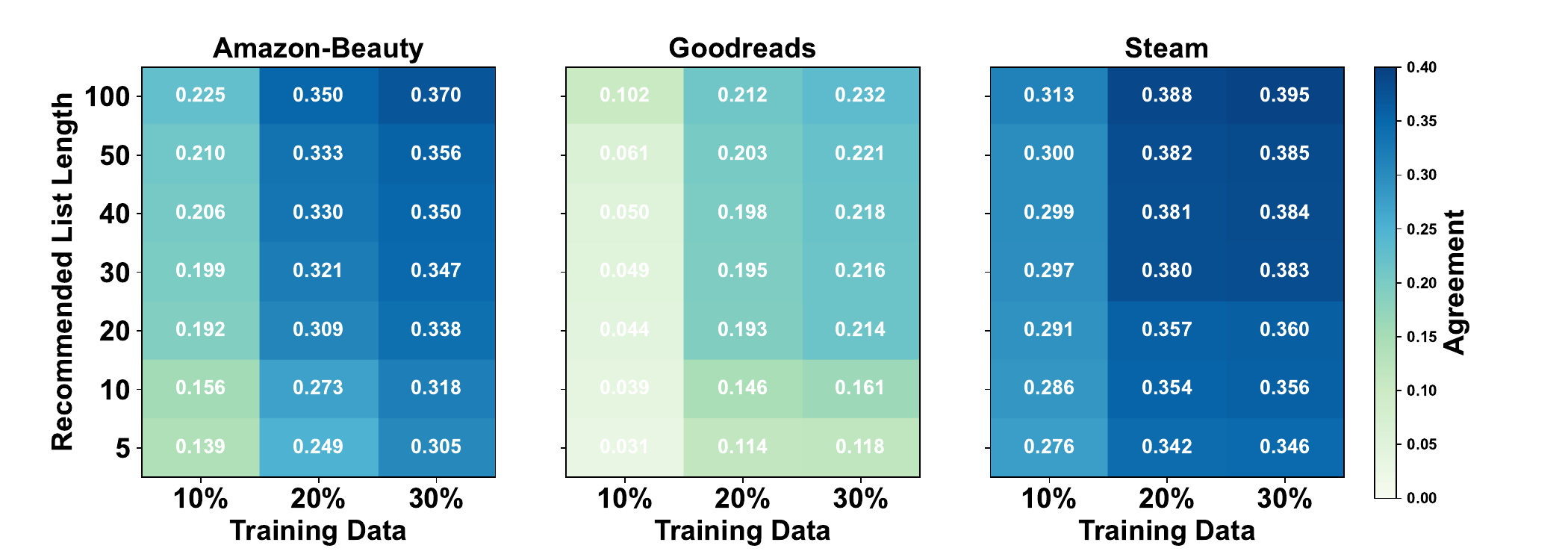}
    \caption{Agreement analysis between surrogate and victim model w.r.t. different recommendation list length across three datasets under different ratio of training data.}
    \label{heatMap}
\end{figure}

To evaluate the effect of training data availability on surrogate model training, we measure the agreement between the surrogate and victim models under varying training ratios (10\%, 20\%, and 30\%). The results are reported across recommendation list lengths ranging from Top-5 to Top-100 on three datasets, as shown in Fig. \ref{heatMap}.

As observed, increasing the training ratio from 10\% to 30\% consistently improves agreement across all datasets, indicating that additional observed interactions help the surrogate better approximate the victim model’s recommendation behavior. Furthermore, agreement increases as the recommendation list length increases. Compared with Top-5, substantially higher values are achieved at Top-50 and Top-100. This suggests that although accurately recovering top-ranked items remains challenging under limited supervision, the surrogate can more reliably capture the victim model’s coarse-grained ranking preferences.


\section{Conclusion}
In this paper, we investigate security vulnerabilities of large language model–powered sequential recommender systems (LLM-SRSs) under realistic black-box settings and propose PUDA, a Prompt-Unknown Dual-Poisoning attack framework that combines evolutionary prompt distillation with coordinated textual and behavioral poisoning. By recovering high-fidelity system instructions and constructing a behavior-aligned surrogate model, PUDA enables effective item promotion without access to proprietary prompts or model internals. Extensive experiments across three real-world datasets show that our method consistently boosts target-item exposure, even under limited attacker budgets. These findings reveal that LLM-SRSs remain highly vulnerable to carefully design attacks.

\begin{acks}
The Australian Research Council partially supports this work under the streams of Future Fellowship (Grant No. FT210100624), the Discovery Project (Grant No. DP240101108, DP260100326 and DP240101814), Discovery Early Career Researcher Award (Grant No. DE230101033 and DE250100613), the Linkage Project (Grant No. LP230200892 and LP240200546), and the Queensland-Bavaria Collaborative Research Program (Grant No. QLDBAVSEED25036).
\end{acks}

\bibliographystyle{ACM-Reference-Format}
\balance
\bibliography{sample-base}

\appendix

\end{document}